\documentclass[prb,aps,twocolumn,showpacs]{revtex4-1}
\pdfoutput=1
\usepackage{dcolumn}
\usepackage{bm}
\usepackage[colorlinks,hyperindex, urlcolor=blue, linkcolor=blue,citecolor=black, linkbordercolor={.7 .8 .8}]{hyperref}
\usepackage{graphicx} 
\usepackage{amsfonts}
\usepackage{amsmath}
\usepackage{amssymb}
\usepackage{amsbsy}
\usepackage{nicefrac}
\usepackage{color,soul}

\begin{document}

\title {Unconventional field-effect transistor composed of electrons floating on liquid helium}

\author{K. Nasyedkin}
\author{H. Byeon}
\author{L. Zhang}
\author{N.R. Beysengulov}
\author{J. Milem}
\author{S. Hemmerle}
\author{R. Loloee} 
\author{J. Pollanen}
\email[]{pollanen@pa.msu.edu}
\affiliation{Department of Physics and Astronomy, Michigan State University, East Lansing, MI 48824-2320, USA}

\date{\today}

\begin{abstract}
We report on an unconventional \emph{macroscopic} field effect transistor composed of electrons floating above the surface of superfluid helium. With this device unique transport regimes are realized in which the charge density of the electron layer can be controlled in a manner not possible in other material systems. In particular, we are able to manipulate the collective behavior of the electrons to produce a highly non-uniform, but precisely controlled, charge density to reveal a negative source-drain current. This behavior can be understood by considering the propagation of damped charge oscillations along a transmission line formed by the inhomogeneous sheet of two-dimensional electrons above, and between, the source and drain electrodes of the transistor.
\end{abstract}

\maketitle

\section{Introduction}
Electrons can be deposited and trapped at a helium vapor-liquid interface to form a extremely high-quality, and exquisitely controllable, two-dimensional electron system (2DES) floating in isolation approximately 10 nm above the helium surface\cite{And97, Mon04}. This 2DES is a paradigm for studying collective effects in strongly-correlated non-degenerate low-dimensional electron systems. While electrical transport measurements provide a window into many of the unique properties of this 2DES, the majority of previous measurements have been performed with a homogeneous areal electron density. However, non-equilibrium phenomena can be realized by developing devices for producing, and precisely controlling, an inhomogeneous spatial distribution of electrons. In fact, creating non-uniform charge devices with electrons on helium has revealed a rich variety of physical phenomena such as the existence of long-lived inter-edge magnetoplasmons\cite{Som95}, dynamic reorganization of the electron density under microwave excitation\cite{Kon12, Bad14, Che15}, and electron crystallization\cite{Ree16, Bad16} and quasi-1D transport\cite{Kli00} under strong spatial confinement. Furthermore, in the limiting case of this type of charge inhomogeneity, where one electron is held in isolation, the toolkit of circuit quantum electrodynamics (cQED) opens the door for novel quantum information experiments with electrons on helium\cite{Pla99, Dyk03, Lyo06, Sch10, Yan16, Abd16}.

In this paper we present the results of electrical transport measurements in a device operated in an unconventional mode that allows us to create, and control, a highly non-uniform electron density. In this mode of operation the device functions as a \emph{macroscopic} field effect transistor (FET) of electrons on helium. Finite element methods (FEM) allow us to simulate the electron density profile for the various regimes of FET operation and compare with our experiment in which we characterize the performance of the FET over a wide frequency range and in a temperature regime where the scattering of electrons from helium vapor atoms produces a strongly temperature dependent conductivity. We find that the imaginary component of the source-to-drain current in the electron layer can become negative at sufficiently high temperature or in the vicinity of FET turn-on where the charge density is maximally inhomogeneous. This seemingly surprising behavior can be explained in terms of a transmission line model between regions of the device with differing electron density.
   
\section{Experiment and FEM modeling}
\begin{figure*}
\begin{center}
\includegraphics[width = \linewidth]{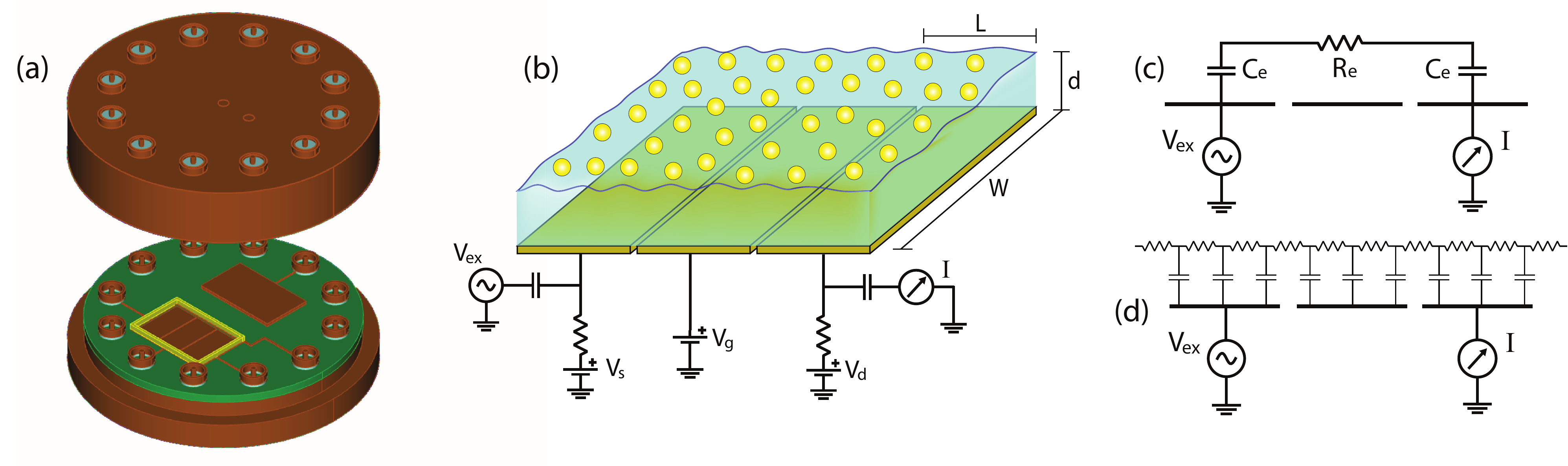}
\end{center}
\caption{(Color online) Schematic of the experimental setup and circuit modeling of electron on helium. (a) 3D CAD rendering of the experimental cell with the array of FET electrodes and a liquid helium level sensor. (b) Sketch of the macroscopic FET device composed of electrons on liquid helium. The dimensions of each electrode are $L \times W=5\times10$~mm and the gap between them is $0.2$~mm. (c) The equivalent lumped RC circuit of the FET device. $R_e$ is the resistance of the electron layer above gate electrode and $C_e$ is the capacitance of electron system to the source and drain electrodes. (d) Transmission line equivalent circuit model of the electron system on liquid helium. In this transmission line mapping, the resistance of the electron layer and its capacitance to the FET electrodes is spatially distributed.}
\label{fig1}
\end{figure*}

Our macroscopic FET of electrons on helium consists of a set of three electrodes located inside of a hermetically sealed experimental cell as shown in Fig.~\ref{fig1}(a). We use a superfluid leak-tight cylindrical copper cell similar to the type developed for recent circuit quantum electrodynamic measurements of electrons on helium\cite{Fra13, Yan16}. The cell is attached a closed-cycle $1$K cryostat for trapping electrons and measuring their low temperature transport properties. Since conventional transport techniques are not possible in this system due to the inability to attach direct Ohmic contacts to the electrons, the electrical properties must be measured via capacitive coupling between the electrons floating on the helium surface and the measuring electrodes submerged beneath. A printed circuit board (PCB) mounted in the cell contains an array of three co-linear rectangular electrodes, which constitute the source, gate and drain of a field effect transistor (FET) device as illustrated in Fig.~\ref{fig1}(b). The FET is surrounded by a negatively biased rectangular guard electrode, which provides lateral electron confinement\footnote{We have tested the experimental results reported in this manuscript with varying values of guard voltage ranging from 0 V to $-50$ V and we find that there is no significant modification to the transport behavior of our device. The effect of a reduced guard voltage is manifest as a slight increase of $V_{th}$ by $\approx 1.5$ V, which is in good agreement with our FEM simulations.}. The device is covered by layer of liquid helium of thickness $d$, which we measure capacitively using a level sensor located adjacent to the FET device. For the measurements reported here the thickness of the liquid helium was $d\simeq0.5$~mm. Thermionic emission from a tungsten filament attached above the FET electrodes provided a source of electrons for our experiments. In addition to the AC voltages used for the transport measurements the FET electrodes are also positively biased with a DC voltage\footnote{\label{note1}We use resistive bias tees consisting of a $1$~$\mu$F capacitors and a $10$~MOhm resistor to apply both AC and DC voltages to a particular FET electrode.} during electron deposition to create an electric field perpendicular to the liquid helium surface to aid in trapping electrons.

In contrast to typical transport experiments with this type of electrode configuration, where electrons are evenly distributed over the all area of the device, we can use differing DC voltages on each electrode to create a spatially inhomogeneous electron density. It is in this way we can operate the device as a FET where the source and drain biases are equal while the gate electrode voltage is varied. In this mode of operation we apply an AC excitation voltage, $V_{ex} = 0.1$ V, to the source electrode and detect the gate tunable AC current on the drain electrode using standard phase-sensitive lock-in techniques. We have carried out these measurements in the frequency range $f=1-100$~kHz and in the temperature range $T=1.35-2.0$~K, which corresponds to the regime where the electron transport mobility is limited by helium vapor atom scattering\cite{Som71, Sai77, Iye80, Lea97}. 

We are able to gain physical insight into our experimental results through the use of finite element method (FEM) modeling of the electrostatic potential experienced by the 2DES and the resulting electron density profile $n_{s}$ on the device. Specifically, we solve Poisson's equation using FEM to find the density distribution of the 2DES, $n_s(x)$, along the length $L$ of the FET electrodes. The electron system is modeled as a charge continuum on the helium surface with effective length $L_e$, which is determined by imposing the condition of electrostatic equilibrium. We determine total number of electrons in the simulation by allowing the electrochemical potential of the charge sheet to change in response to the gate voltage. Their areal density distribution is then given by $n_s(x) = -(\epsilon_0 / e) (E_{+}(x) - \epsilon E_{-}(x))$, where $E_{+,-}(x)$ are electrical field distributions above and below the charge sheet,$\epsilon_0$ is the vacuum permittivity and $\epsilon = 1.057$ is the dielectric constant of helium. 

\section{Results and Discussion}
\subsection{Electrons on helium FET I-V characteristics}
\begin{figure}
\begin{center}
\includegraphics[width=1 \columnwidth]{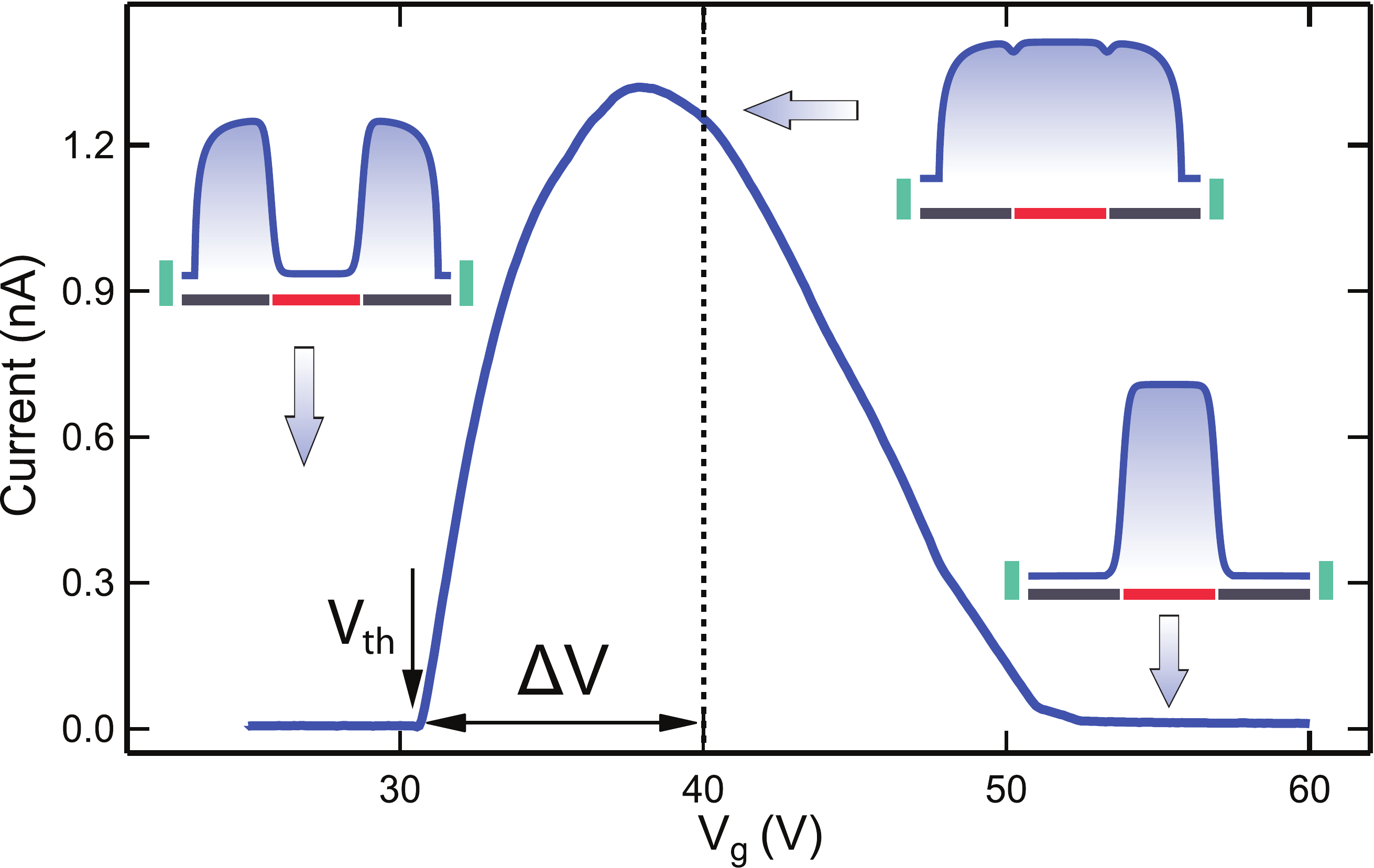}%
\caption{(Color online) Amplitude of the source-to-drain current, $\vert I \vert$, as a function of gate voltage at $T=1.35$~K for the electrons on helium FET device. For these measurements $V_s=V_d=40$~V and the guard electrode was biased with $-50$ V. The insets illustrate the electron charge profile over the FET electrodes obtained by finite element simulation of the device. A uniform electron density is achieved when the source, drain and gate electrodes have equal value (dashed vertical line).}
\label{fig2}
\end{center}
\end{figure}

A typical source-drain current-voltage (I-V) characteristic is shown in Fig.~\ref{fig2} for FET operation. For small values of the gate voltage, $V_g$, no current flows through electron layer because all of the electrons are localized above the source and drain electrodes and the area over the gate is depleted. This is illustrated in left inset of Fig.~\ref{fig2} where we show the electron density profile simulated using FEM. Upon increasing $V_g$ electrons are attracted to the region above the gate leading to the onset of source-to-drain current flow at a threshold value of the gate voltage, $V_{th}$. This behavior mimics that of a conventional semiconductor FET. However, further increasing $V_{g}$ reveals the first of several phenomena unique to a FET of electrons floating on helium. Unlike a conventional FET, the source-drain current reaches a maximum in the vicinity of uniform areal electron density ($V_{s} = V_{d} = V_{g}$), as illustrated by the central inset. After reaching a maximum value, the source-drain current begins to decrease and eventually vanishes with increasing $V_{g}$. This effect results from the reduced mobility of the electrons\cite{Sai77} and their redistribution by the increasing gate field. For sufficiently large $V_{g}$ all of the electrons will be located above the gate, leading to $|I|=0$, since the charge sheet is not connected to a ground reservoir of electrons, but rather operates with a fixed number of particles. This effect is highlighted in the right inset of Fig.~\ref{fig2} where we show the FEM simulated density profile for large gate voltage.

While this behavior is markedly different from FET devices fabricated in other material systems having non-fixed electron number, the entire I-V curve shown Fig.~\ref{fig2} can, however, be understood in terms of the equivalent lumped RC circuit model shown in Fig.~\ref{fig1}(c), where the electron layer with resistance $R_e$ is coupled to the source and drain electrodes by capacitance $C_e$.

We can estimate the number of electrons above the FET by considering the capacitive redistribution of electrons to the region above the gate electrode during a gate sweep. In particular, the difference $\Delta V = V_{s,d}-V_{th}$ between the threshold (turn-on) voltage $V_{th}$ and the voltages that lead to an equal number of electrons over each electrode, \emph{i.e.}~$V_s=V_g=V_d$, is related to the number of electrons above the FET device by\cite{Fra13} 
\begin{equation}
\label{N_e}
N = 2WL\Delta V\frac{\epsilon\epsilon_0}{ed}
\end{equation}
For the measurements shown in Fig.~\ref{fig2}, where $V_s=V_d=40$~V, the total number of electrons above the FET device in this case is approximately $1\times10^8$, corresponding to an areal density of $n=7\times10^7$~cm$^{-2}$ for our electrode dimensions and the case when the electrons are evenly distributed over the three FET electrodes, $V_s=V_d=V_g$. This electron density is also consistent with that obtained by FEM simulations.

\begin{figure}
\begin{center}
\includegraphics[width=1 \columnwidth]{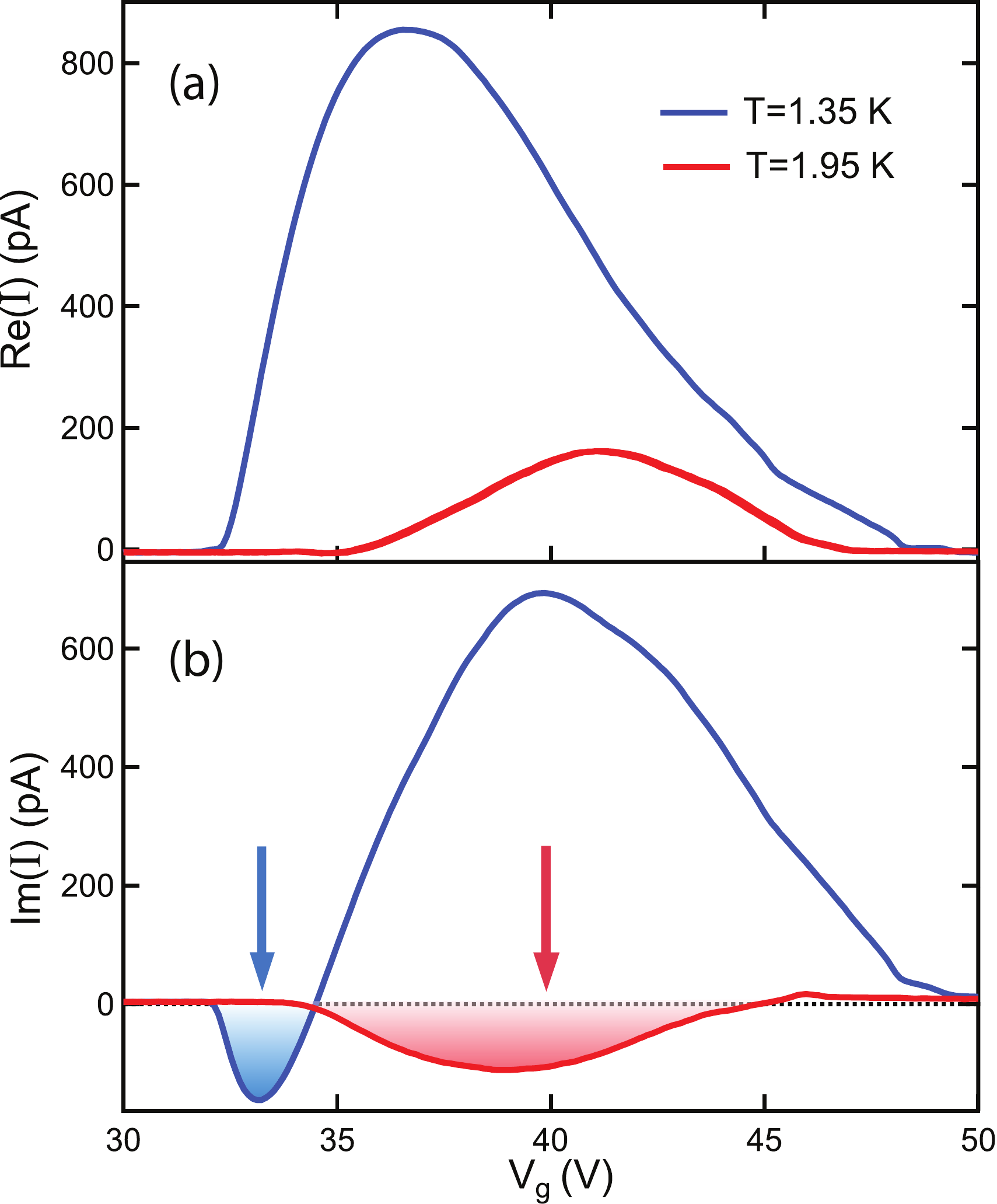}
\caption{(Color online) Real (a) and imaginary (b) components of the FET source-drain current as a function of gate voltage measured at $T=1.35$~K (blue curve) and $T=1.95$~K (red curve) at $f = 60$~KHz. For these measurements $V_s=V_d=40$~V and the guard electrode was biased with $-50$ V.} The negative values of $\operatorname{Im}(I)$, indicated by the vertical arrows and shaded regions, represent a unique departure from the lumped RC circuit model for an FET made of electrons on helium.
\label{fig3}
\end{center}
\end{figure}

Lock-in measurements allow us to simultaneously measure the in-phase (real) and quadrature (imaginary) components $\operatorname{Re}(I)$ and $\operatorname{Im}(I)$ of the complex source drain current. Knowledge of both components is necessary to accurately model the impedance of the system as a function of frequency since it contains both resistive and reactive elements. In Fig.~\ref{fig3} we show the gate voltage dependence of the real (a) and imaginary (b) components of the source-drain current measured at $T=1.35$~K (blue trace) and $T=1.95$~K (red trace). While at low temperature both $\operatorname{Re}(I)$ and $\operatorname{Im}(I)$ are positive as expected from the lumped circuit model, in contrast we find that at high temperature the data exhibit an anomalous gate voltage dependence where the imaginary component of the current $\operatorname{Im}(I)$ is negative at sufficiently high temperature (shaded red region). This negative current implies that the relative phase angle $\phi$ between the source-drain current and the AC excitation voltage is also negative at high temperature. Furthermore, negative values of $\operatorname{Im}(I)$ are not restricted only to high temperature, but rather also manifest in the vicinity of FET depletion when $V_g \gtrsim V_{th}$ shown as the shaded blue region in Fig.~\ref{fig3}(b)\footnote{The small shift in turn-on voltage between high and low temperature is due to a change in the helium layer thickness induced by increasing the temperature. This shift is consistent with change in helium thickness we measure with our \emph{in-situ} capacitive level sensor.}.

\subsection{Transmission line mapping}
This negative value of the quadrature source-drain current cannot be explained by the lumped RC model for AC transport of electrons on helium for which both $\operatorname{Re}(I)$ and $\operatorname{Im}(I)$ are strictly positive for all values of $R_e$ and $C_e$. To understand values of $\operatorname{Im}(I) < 0$ the wave nature of the propagating electrical signal must be taken into account and the 2DES on helium should be considered as a transmission line with spatially distributed resistance and capacitance (Fig.~\ref{fig1}(d)). The need to describe the AC transport of electrons on helium in this fashion was first pointed out by Sommer and Tanner\cite{Som71}. For a homogeneous distribution of electrons in zero magnetic field this analysis was developed by Mehrotra and Dahm\cite{Meh87} and by Lea \emph{et al.}\cite{Lea91} for both zero-field and magneto-transport of electrons on helium.

We apply a similar transmission line mapping and analysis to of our macroscopic FET. For the rectangular geometry of our device we assume no variation in the voltage and current along the width $W$ of the electrodes (see Fig.~\ref{fig1}(b)), therefore modeling of the system reduces to a one-dimensional RC transmission line, where a damped voltage wave propagates from the source to the drain with the boundary condition that the current density is zero at both of these electrodes. We can calculate the complex current $I=\operatorname{Re}(I) + j\operatorname{Im}(I)$ produced by this voltage wave by following an analysis similar to that developed by Lea \emph{et al.}\cite{Lea91} and Mehrotra and Dahm\cite{Meh87}
\begin{equation}
\label{I*}
\frac{I}{I_0} = (1+j) \frac{3\delta}{2L} \frac{\sinh^2(jkL)}{\sinh(3jkL)},
\end{equation}
where
\begin{equation}
\label{I0}
I_0 = \omega C'(WL/3)V_{ex},
\end{equation}
is a normalization current\cite{Lea91},
\begin{equation}
\label{k}
k = \frac{1-j}{\delta},
\end{equation}
is the complex wavevector of the damped voltage wave,
\begin{equation}
\label{delta0}
\delta = \sqrt{\frac{2\sigma}{\omega C'}},
\end{equation}
is a characteristic decay length over which charge density fluctuation propagates in the electron layer,
\begin{equation}
\label{C'}
C' = \epsilon_0\left(\frac{\epsilon}{d}+\frac{1}{D-d}\right),
\end{equation}
is the capacitance between the electron layer and the top\footnote{Here the top electrode is formed by the interior portion of the experimental cell body directly above the electron layer and FET electrodes.} and bottom electrodes, which are spaced by a distance $D$, and
\begin{equation}
\label{sigma}
\sigma = ne\mu
\end{equation}
is the Drude conductivity of the 2DES. Here $\mu$ is the mobility of the 2DES and $\omega=2\pi f$ is the angular excitation frequency driving electrical transport through the device. 

\begin{figure}[b]
\begin{center}
\includegraphics[width=1 \columnwidth]{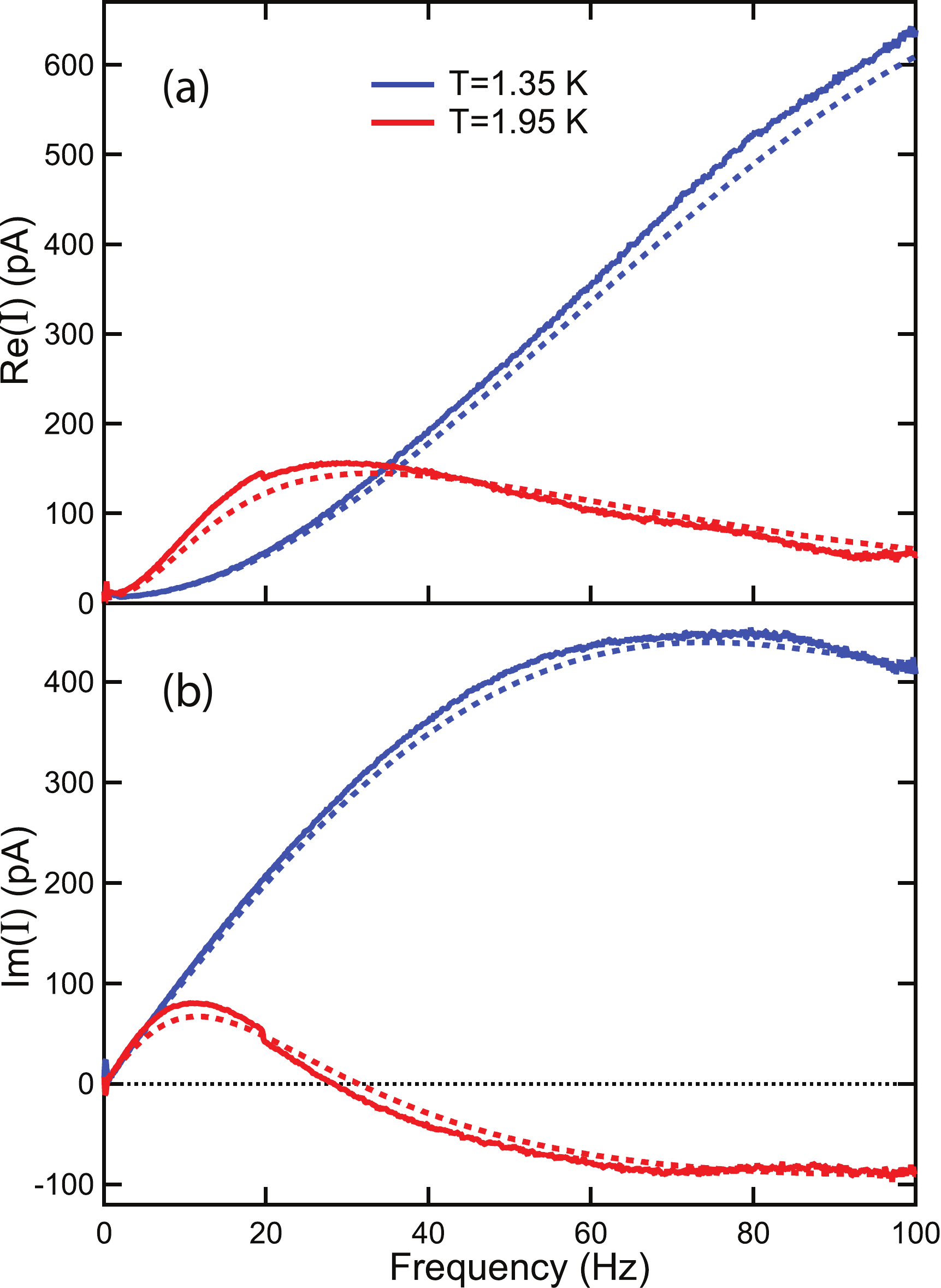}
\caption{(Color online) Frequency dependence of the real $\operatorname{Re}(I)$ (a) and imaginary $\operatorname{Im}(I)$ (b) components of the source-drain current with a uniform electron density above all three FET electrodes corresponding to $V_s=V_d=V_g=40$~V. For these measurements the guard electrode was biased with $-5$ V. The solid traces show the experimental data obtained at $T=1.35$~K (blue) and $T=1.95$~K (red). The correspondingly colored dashed lines are calculated using the transmission line model described in the text.}
\label{fig4}
\end{center}
\end{figure}
Eq. (\ref{I*}) provides a qualitative understanding of the data depicted in Fig~\ref{fig2}. For our electrode geometry the normalized current depends only on the parameter $\delta$, which can be regarded as a two-dimensional AC skin depth\cite{Lea90} analogous to the AC electric field penetration length near the surface of three-dimensional conductors. When $\delta$ is comparable or smaller than the total length of the FET electrodes, $\delta \lesssim 3L$, the imaginary component of the source-drain current reverses sign since the phase angle between the current and driving voltage $\phi=\frac{\pi}{2}-3kL$ becomes negative. At low temperature the electron mobility, and corresponding conductivity, are relatively high. As a result $\delta$ is larger than the overall length of the FET electrodes and $\operatorname{Re}(I)$ and $\operatorname{Im}(I)$ remain positive. Increasing the temperature leads to an exponential increase of the density of helium vapor atoms and hence to a marked decrease of the electron mobility. In the range of temperatures $T=1.35-1.95$~K the reduction of mobility is approximately an order of the magnitude, which reduces $\delta$ to a value much smaller than the total length of the electrodes leading to the negative values of $\operatorname{Im}(I)$ we observe in our measurement. Furthermore, our finding that $\operatorname{Im}(I)<0$ at low temperature ($T=1.35$~K) but only in the vicinity of FET turn-on can be understood in a similar fashion. When $V_g \simeq V_{th}$ the number of electrons present in the gate region is small and according to Eq.~\ref{sigma} their resistivity is relatively high leading to a decrease in $\delta$ and $\operatorname{Im}(I)<0$. As $V_g$ is further increased more electrons are attracted to the region above the gate electrode, which increases the 2DES conductivity and tunes the sign of $\operatorname{Im}(I)$ to positive values.

We have also performed a quantitative analysis based on the transmission line model for the case where the electron density is homogeneously distributed over the device. This model is defined by Eq.~\ref{I*}-\ref{sigma}, which relate the mobility $\mu$ to the frequency dependent complex source-drain current. The comparison between a fit based on this model (dashed curves), with the transport mobility $\mu$ as a fitting parameter, and the experimental data (solid curves) is shown in Fig~\ref{fig4}, where the frequency dependence of both components of the measured current is plotted at low ($T=1.35$~K) and high ($T=1.95$~K) temperatures. For the data in Fig~\ref{fig4} the DC bias potential is the same for all FET electrodes $V_s=V_d=V_g=40$~V and the electron density is $n=7 \times 10^7$~cm$^{-2}$. The value of the mobility obtained from this fitting is $2.8 \times 10^{4}$ cm$^2$/Vs at $T=1.35$ K and $0.45 \times 10^{4}$ cm$^2$/Vs at $T=1.95$ K.  These values are in reasonable agreement with previous measurements\cite{Iye80} and the theoretical values given by Saitoh\cite{Sai77}.

Finally we note that this type of transmission line mapping applies not only to electrons on helium but is also needed to describe a variety of phenomena in quantum \emph{degenerate} 2DES such as measurements of the two-dimensional metal-insulator transition in semiconductor heterostructures\cite{Dul00,Tra06} and the propagation of plasmons in graphene striplines\cite{Gom13}. However the low level of disorder and high degree of controllability provided by the system of electrons on helium allows us to explore the ramification of this model in a regime of extreme charge inhomogeneity difficult to replicate in other material systems.

\section{Conclusion}
In summary, we have investigated the operation of a macroscopic field effect transistor composed of electrons on liquid helium in a regime where electron scattering from helium vapor atoms is the dominant form of disorder. In this FET mode of operation the electron density can be made highly non-uniform and we find that the imaginary component of the source-drain current changes sign from positive to negative when the systems is subjected to sufficiently high temperatures or when the electron density is made maximally inhomogeneous. These transport regimes can be understood by considering the system of electrons as a voltage wave propagating in a transmission line composed of the electrons floating above the metallic electrodes under the helium surface. We anticipate that an FET mode of operation could find application in studying other non-equilibrium phenomena such as ultrahot electron on helium\cite{Kon07, Nas09, Che18, Kle18}, electrons strongly confined in helium microchannel devices or for future high frequency surface acoustic wave experiments 

\begin{acknowledgements}
We are grateful to M.I. Dykman, J.R. Lane, D.G. Rees, D.I. Schuster, G. Koolstra, S.S. Sokolov, H. Choi, K. Kim, W.P. Halperin and N.O. Birge for helpful discussion. This work was supported by the NSF (Grant no. DMR-1708331).
\end{acknowledgements}

\bibliographystyle{apsrev4-1} 

\end{document}